\documentclass[
reprint,
showpacs,
amsmath,amssymb,aps,
pra,
]{revtex4-1}

\usepackage{graphicx}
\usepackage{dcolumn}
\usepackage{bm}
\usepackage{hyperref}
\usepackage[mathlines]{lineno}
\usepackage{xcolor}
\usepackage{braket}

\begin{document}

\preprint{}

\title{Parametric amplifier for Bell measurement in continuous-variable quantum state teleportation}

\author{Xin Chen}
\author{Z. Y. Ou}
\email{zou@iupui.edu}
\affiliation{%
Department of Physics, Indiana University-Purdue University Indianapolis, Indianapolis, IN 46202, USA
}%





\begin{abstract}
A parametric amplifier is in essence a linear four-port device, which couples and linearly mixes two inputs  before amplifying and sending them to two output ports. Here, we show that for quadrature-phase amplitudes, a parametric amplifier can replace beam splitters to play the role of mixer. We apply this idea to a continuous-variable quantum state teleportation scheme in which a parametric amplifier replaces a beam splitter in Bell measurement. We show that this scheme is loss-tolerant in the Bell measurement process and thus demonstrate the advantage of PA over BS in the applications in quantum measurement.
\end{abstract}


\maketitle

\section{Introduction}

Quantum interference plays an important role in the display of many quantum phenomena. It usually requires a linear beam splitter to superimpose two fields for interference between them. This happens in many protocols of quantum information processing. For example, optical quantum computing relies on the Hong-Ou-Mandel effect where a beam splitter is a part \cite{KLM,oc}. Current applicable schemes of Bell measurement \cite{brau1,brau2} for quantum state teleportation \cite{tele,vaid,bou,fur} require beam splitters to mix the incoming unknown state with one field of an entangled state.

It is well-known that losses are notorious in degrading quantum effects and are the key obstacle in many protocols of quantum information processing. Detection process often introduces losses due to imperfect coupling and less-than-unit quantum efficiency. But highly efficient detectors are only available for some limited spectrum of the electromagnetic waves. Thus, it becomes a major concern in high fidelity quantum communication involving quantum measurement by detection. Quantum state teleportation is one of such quantum communication protocols where a Bell measurement is performed to projectively select out the required states. For continuous-variable quantum state teleportation, Bell measurement is usually achieved by homodyne detection, which is sensitive to losses. This will inevitably affect the fidelity of the teleported state.

On the other hand, amplification is known to overcome the effect of losses. Indeed, parametric amplifiers were recently used in SU(1,1) interferometers \cite{ou12} and quantum entanglement measurement \cite{jml19} to mix two fields in place of beam splitters for interference and was demonstrated to be loss tolerance in detection processes.

At first look, it seems counter-intuitive that a parametric amplifier can be of any use in quantum information science and play any role in mixing fields for interference since it is often portrayed as adding extra noise and thus degrading the signal in the amplification processes \cite{caves}. Nevertheless, if we look into the origin of the extra noise, we find it comes from the vacuum fluctuations of the internal degrees of the amplifier. So, if we can access to these internal degrees and place them in correlation with the input, the extra noise can actually be suppressed due to quantum correlation \cite{ou93,kong13}. Therefore, by treating the internal degrees of the amplifier as another input, we mix it with the original input and can use the amplifier as a field mixer similar to a beam splitter. Specifically, parametric amplifiers are such devices for which the internal degree is the so-called idler field that we can easily access from outside. In essence, a parametric amplifier is a four-port linear device just like a beam splitter, even though it is often realized through nonlinear interaction with energy actively pumped into it for amplification.

In this paper, we will investigate the feasibility of replacing  a beam splitter by a parametric amplifier for Bell measurement in quantum teleportation scheme and demonstrate the loss tolerant property of the new scheme.
The paper is organized as follows. In Sect.II, we introduce quantum state transformation for both beam splitter and parametric amplifier. This is based on Wigner representation of the quantum state. In Sect.III, we present the result for quantum teleportation with a parametric amplifier and demonstrate its feasibility. The tolerant property of the new scheme will be discussed in general in Sect.IV and on the transmission of pure states such as a coherent state and photon number Fock states in Sect.V and on EPR-entangled states in Sect.VI. We conclude with a discussion in Sect.VII.

\section{Quantum state transformation of a parametric amplifier}

The role played a parametric amplifier in the mixing of fields for interference can be understood from the following input-output relation:
\begin{equation}
\label{eq:in-out}
\hat a_{1(PA)}^{(o)} = G \hat a_1^{(i)} + g \hat a_2^{(i)\dag},~~\hat a_{2(PA)}^{(o)} = G \hat a_2^{(i)} + g \hat a_1^{(i)\dag},
\end{equation}
where $G, g$ are amplitude gains satisfying $G^2-g^2=1$ and without loss of generality, we assume they are real and positive.  In comparison, the input-output relation for a beam splitter is given by
\begin{equation}
\label{BS:in-out}
\hat a_{1(BS)}^{(o)} = t \hat a_1^{(i)} + r \hat a_2^{(i)},~~\hat a_{2(BS)}^{(o)} = t \hat a_2^{(i)} - r \hat a_1^{(i)}
\end{equation}
with $t^2+r^2 =1$.
It can be seen that the PA is basically a four-port linear device that not only amplifies but also, similar to a beam splitter, mixes the two input fields.

However, the difference between the two devices is also obvious: the PA output is related to the Hermitian conjugate of the second input field, which can lead to unwanted spontaneous emission even with input in vacuum, as seen in the average photon number:
\begin{eqnarray}
\label{N:in-out}
\langle\hat N_{1(PA)}^{(o)}\rangle &\equiv & \langle \hat a_1^{(o)\dag}\hat a_1^{(o)}\rangle  \cr &=&  G^2 \langle \hat a_1^{(i)\dag}\hat a_1^{(i)}\rangle + g^2 ( \langle \hat a_2^{(i)\dag}\hat a_2^{(i)}\rangle +1),
\end{eqnarray}
if the two inputs are independent of each other. So, PA is not suitable for mixing photons or discrete variable quantum information processing. On the other hand, the input-output relations for quadrature-phase amplitudes are given by
\begin{equation}
\label{PA:XYin-out}
\hat X_{1,2(PA)}^{(o)} = G \hat X_{1,2}^{(i)} + g \hat X_{2,1}^{(i)},~~\hat Y_{1,2(PA)}^{(o)} = G \hat Y_{1,2}^{(i)} - g \hat Y_{2,1}^{(i)},
\end{equation}
which is similar to those for a beam splitter:
\begin{equation}
\label{BS:XYin-out}
\hat X_{1,2(BS)}^{(o)} = t \hat X_{1,2}^{(i)} \pm r \hat X_{2,1}^{(i)},~~\hat Y_{1,2(BS)}^{(o)} = t \hat Y_{1,2}^{(i)} \pm r \hat Y_{2,1}^{(i)},
\end{equation}
where $\hat X\equiv \hat a +\hat a^{\dag}, \hat Y\equiv (\hat a -\hat a^{\dag})/j (j\equiv\sqrt{-1})$ for the corresponding field described by $\hat a$. So, they only differ in coupling coefficients. Therefore, for continuous variable quantum information processing, a PA can play the same role as a BS for superimposing two fields. Note from Eqs.(\ref{PA:XYin-out},\ref{BS:XYin-out}) that similar to the situation of loss, which introduces quantum noise through vacuum in the unused port, amplification also adds noise through the vacuum of the second input if it is unattended and thus uncorrelated with the signal input. For this reason, the second input is usually called ``idler".

The relationships in Eqs.(\ref{PA:XYin-out},\ref{BS:XYin-out}) provide us a way to evaluate quantum state transfer through a BS and a PA, which can be done through the Wigner function.
For a PA with input state described by a Wigner function $W_{in}(X_1,Y_1;X_2,Y_2)$, we can find the output state by considering the characteristic function of two-mode Wigner function:
\begin{eqnarray}
&&\chi(u_1,v_1;u_2,v_2) 
\cr &&\hskip 0.5in =\textrm{Tr}(\hat{\rho}e^{jv_1\hat X_1-ju_1\hat Y_1+jv_2\hat X_2-ju_2\hat Y_2})
\cr &&\hskip 0.5in =\int dx_1 dy_1dx_2dy_2 ~~W(x_1,y_1;x_2,y_2)\cr&&\hskip 0.9in \times  e^{ jv_1x_1-ju_1y_1+jv_2x_2-ju_2y_2},\label{chi}
\end{eqnarray}
Since the input-output relations presented in Eqs.(\ref{PA:XYin-out},\ref{BS:XYin-out}) are for Heisenberg picture, the state described by density operator $\hat \rho$ is the same for both input and output. Using Eq.(\ref{PA:XYin-out}), we find
\begin{eqnarray}
&&\chi_{out}(u_1,v_1;u_2,v_2) \cr && \hskip 0.5in =\textrm{Tr}(\hat{\rho}e^{jv_1\hat X_1^{(o)}-ju_1\hat Y_1^{(o)}+jv_2\hat X_2^{(o)}-ju_2\hat Y_2^{(o)}})
\cr && \hskip 0.5in =\textrm{Tr}(\hat{\rho}e^{jv_1'\hat X_1^{(i)}-ju_1'\hat Y_1^{(i)}+jv_2'\hat X_2^{(i)}-ju_2'\hat Y_2^{(i)}})
\cr && \hskip 0.5in =C_{W_{in}}(v_1',u_1';v_2',u_2'),
\end{eqnarray}
where $u_1'=u_1G-u_2g$, $v_1'=v_1G+v_2g$, $u_2'=u_2G-u_1g$, $v_2'=v_2G+v_1g$. Taking reverse Fourier transformation for $W$, we find
\begin{eqnarray}
\label{PA:Win-out}
&&W_{out}^{(PA)}(x_1,y_1;x_2,y_2) \cr
&&= W_{in}(Gx_1-gx_2,Gy_1+gy_2;Gx_2-gx_1,Gy_2+gy_1).\cr &&
\end{eqnarray}
Similarly for a BS with the same input state,  the output state is described by \cite{ou17}
\begin{eqnarray}
\label{BS:Win-out}
&&W_{out}^{(BS)}(x_1,y_1;x_2,y_2) \cr
&&= W_{in}(tx_1-rx_2,ty_1-ry_2;tx_2+rx_1,ty_2+ry_1).~~~~
\end{eqnarray}
Comparing Eqs.(\ref{PA:Win-out},\ref{BS:Win-out}), we find the output Wigner functions for the two devices give rise to superposition of input fields but with different phases and different transfer coefficients.

As an example, let us consider the input of a two-mode squeezed state with a Wigner function of \cite{ou93}
\begin{eqnarray}\label{EPR}
&&W_{in}(x_1,y_1;x_2,y_2)\cr &&\hskip 0.3 in =\frac{1}{(2\pi)^2}e^{-\frac{1}{4}[(x_1+x_2)^2+(y_1-y_2)^2]e^{2s}}\cr &&\hskip 0.9 in  \times e^{-\frac{1}{4}[(x_1-x_2)^2+(y_1+y_2)^2]e^{-2s}},
\end{eqnarray}
where $s$ is the squeezing parameter. It is known that when $t=r=1/\sqrt{2}$, the output of BS is two single-mode squeezed states with squeezing at orthogonal quadratures. This can be easily confirmed from Eq.(\ref{BS:Win-out}):
\begin{eqnarray}
\label{BS:Win-out-sq}
&&W_{out}^{(BS)}(x_1,y_1;x_2,y_2) \cr &&\hskip 0.3 in=\frac{1}{(2\pi)^2}e^{-\frac{1}{2}(x_1^2+y_2^2)e^{2s}}  e^{-\frac{1}{2}(x_2^2+y_1^2)e^{-2s}}
\cr &&\hskip 0.3 in =\frac{1}{2\pi}e^{-\frac{1}{2}(x_1^2e^{2s}+y_1^2e^{-2s})} \frac{1}{2\pi}e^{-\frac{1}{2}(x_2^2e^{-2s}+y_2^2e^{2s})}.~~~~
\end{eqnarray}
The corresponding situation for a PA is:
\begin{eqnarray}
\label{PA:Win-out-sq}
&&W_{out}^{(PA)}(x_1,y_1;x_2,y_2) \cr &&\hskip 0.3 in= \frac{1}{(2\pi)^2}e^{-\frac{1}{4}[(x_1+x_2)^2+(y_1-y_2)^2]e^{2s}(G-g)^2}  \cr &&\hskip 0.7 in\times e^{-\frac{1}{4}[(x_1-x_2)^2+(y_1+y_2)^2)]e^{-2s}(G+g)^2}.~~~~
\end{eqnarray}
Especially when $G+g =1/(G-g) = e^{s}$, we have
\begin{eqnarray}
\label{PA:Win-out-sq2}
&&W_{out}^{(PA)}(x_1,y_1;x_2,y_2)
\cr &&\hskip 0.3 in =\frac{1}{2\pi}e^{-\frac{1}{2}(x_1^2+y_1^2)} \frac{1}{2\pi}e^{-\frac{1}{2}(x_2^2+y_2^2)},~~~~
\end{eqnarray}
which is just the Wigner function for vacuum. This is equivalent to the case of balanced gain in an SU(1,1) interferometer \cite{ou12}.

\begin{figure*}
\centering
\includegraphics[width=14cm]{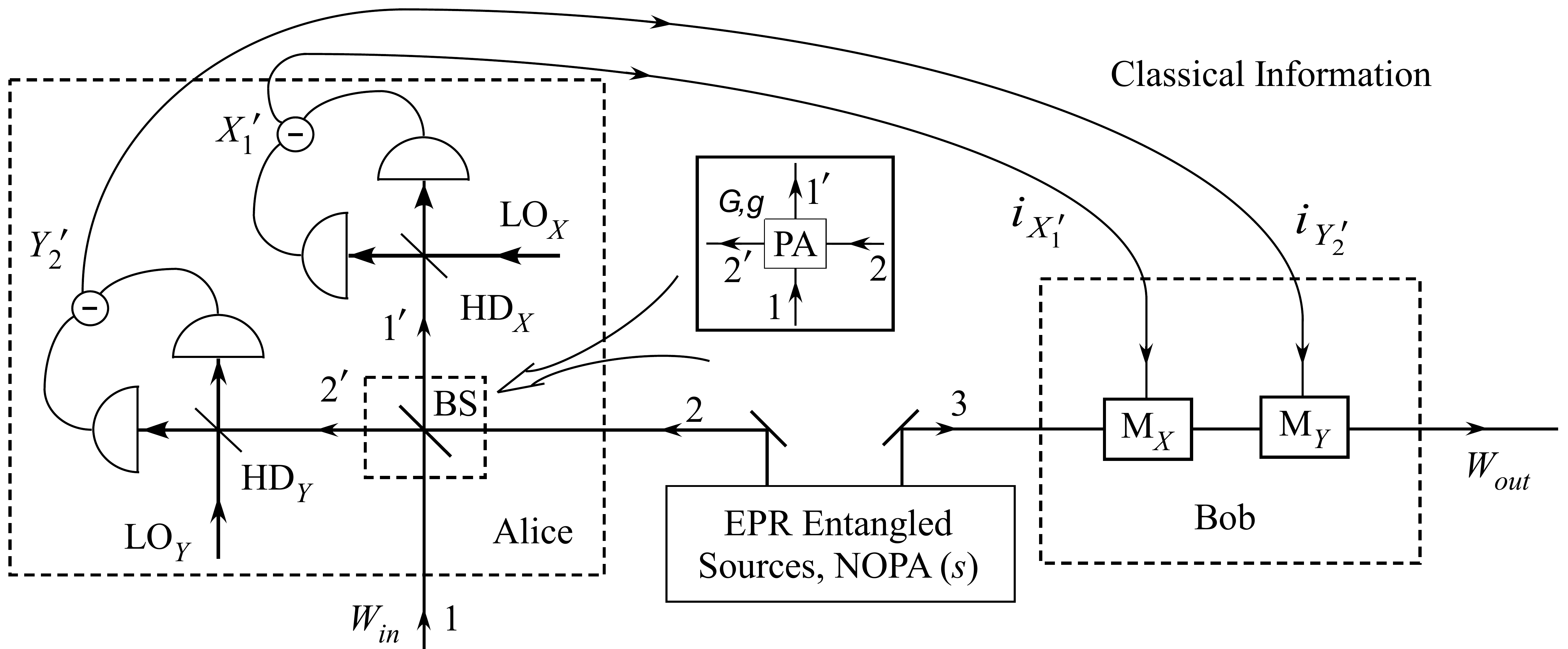}
\caption{Schematics for the continuous-variable quantum state teleportation with a parametric amplifier (PA, inset) in place of a beam splitter (BS).
}
\label{fig:ill}
\end{figure*}

\section{Application to quantum state teleportation}

Consider the scheme shown in Fig.\ref{fig:ill} for  quantum state teleportation with continuous variables. This is the scheme that makes a complete state teleportation (even for vacuum state) \cite{brau2,fur}. In this scheme, Bell projection measurement is performed with a beam splitter (BS) to mix the unknown state to be teleported with one of the field of the EPR-entangled state, which is described in Eq.(\ref{EPR}). It has been shown that parametric amplifier-assisted homodyne measurement can make the same quantum correlation measurement as the homodyne measurement \cite{jml19}. So, let us now replace the BS with a parametric amplifier (PA) of gain parameters $G,g$, shown in the inset of Fig.\ref{fig:ill}. Using Eq.(\ref{EPR}) for the EPR entangled state with a strength of $s$ and labeling of the fields in Fig.\ref{fig:ill}, we find the input state
of the parametric amplifier is described by the Wigner function:
\begin{eqnarray}\label{W-in}
&&W_{in}(x_1,y_1;x_2,y_2;x_3,y_3)\cr &&\hskip 0.2 in =\frac{1}{(2\pi)^2}e^{-\frac{1}{4}[(x_3+x_2)^2+(y_3-y_2)^2]e^{2s}}\cr &&\hskip 0.4 in  \times e^{-\frac{1}{4}[(x_3-x_2)^2+(y_3+y_2)^2]e^{-2s}} W_{in}(x_1,y_1).
\end{eqnarray}
From Eq.(\ref{PA:Win-out}), the Wigner function after the PA becomes
\begin{eqnarray}\label{W-PA}
&&W_{out}(x'_1,y'_1;x'_2,y'_2;x_3,y_3)\cr &&\hskip 0.2 in = \frac{1}{(2\pi)^2}e^{-\frac{1}{4}[(x_3+Gx'_2-gx'_1)^2+(y_3-gy'_1-Gy'_2)^2]e^{2s}}\cr &&\hskip 0.5 in\times e^{-\frac{1}{4}[(x_3-Gx'_2+gx'_1)^2+(y_3+gy'_1+Gy'_2)^2]e^{-2s}}
\cr &&\hskip 0.7 in\times W_{in}(Gx'_1-gx'_2,Gy'_1+gy'_2).
\end{eqnarray}

Now we make homodyne measurement of $\hat X_1$ and $\hat Y_2$ of the PA output fields. With a result of $i_{X_1'}$ and $i_{Y_2'}$, the state in field 3 (other field of the EPR-entangled state) is projected to a state described by the density operator:
\begin{eqnarray}
\hat \rho_{proj} = {\rm Tr}_{1'2'}(|i_{X_1'},i_{Y_2'}\rangle\langle i_{X_1'},i_{Y_2'}|\hat \rho_{sys}),
\end{eqnarray}
where $|i_{X_1'},i_{Y_2'}\rangle$ is the common eigen state of $\hat X_1$ and $\hat Y_2$.
The Wigner function of the projected state is then
\begin{eqnarray}
&&W_{proj}(x_3,y_3)\cr &&\hskip 0.2 in=\int dx_2'dy_1' W_{out}(x'_1,y'_1;x'_2,y'_2;x_3,y_3)|_{x_1'=i_{X_1'},y_2'=i_{Y_2'}}
\cr &&\hskip 0.2 in=\int \frac{dx_2'dy_1'}{(2\pi)^2}e^{-\frac{1}{4}[(x_3+Gx'_2-gi_{X_1'})^2+(y_3-gy'_1-Gi_{Y_2'})^2]e^{2s}}
\cr &&\hskip 0.5 in\times e^{-\frac{1}{4}[(x_3-Gx'_2+gi_{X_1'})^2+(y_3+gy'_1+Gi_{Y_2'})^2]e^{-2s}}
\cr &&\hskip 0.7 in\times W_{in}(Gi_{X_1'}-gx'_2,Gy'_1+gi_{Y_2'})
\end{eqnarray}
Combining the common terms in the integral above, we obtain
\begin{eqnarray}
&&W_{proj}(x_3,y_3)\cr &&\hskip 0.2 in =\frac{e^{-\frac{x_3^2+y_3^2}{2\textrm{cosh}2s}}}{(2\pi)^2}
\int dx_2'dy_1'e^{-\frac{\textrm{cosh}2s}{2}(Gx'_2+x_3\textrm{tanh}2s-gi_{X_1'})^{2}}
\cr &&\hskip 0.8 in\times e^{-\frac{\textrm{cosh}2s}{2}(gy'_1+Gi_{Y_2'}-y_3\textrm{tanh}2s)^2}
\cr &&\hskip 1 in\times W_{in}(Gi_{X_1'}-gx'_2,Gy'_1+gi_{Y_2'}).~~
\end{eqnarray}
Taking the limit of $s\gg1$ so that $\cosh 2s \gg 1$ or the range of $W_{i}(x_3,y_3)$, we can approximate the Gaussian functions in the integral above with $\delta$-functions. Then we have
\begin{eqnarray}
&&W_{proj}(x_3,y_3)
\cr &&\hskip 0.2 in=\frac{1}{2\pi\textrm{cosh}2s}e^{-\frac{1}{2\textrm{cosh}2s}(x_3^2+y_3^2)}
\cr &&\hskip 0.5 in\times \int dx_2'dy_1'\delta(Gx'_2+x_3\textrm{tanh}2s-gi_{X'_1})
\cr &&\hskip 0.7 in\times \delta(gy'_1+Gi_{Y_2'}-y_3\textrm{tanh}2s)
\cr &&\hskip 0.9 in\times W_{in}(Gi_{X_1'}-gx'_2,Gy'_1+gi_{Y_2'})
\cr &&\hskip 0.2 in
=\frac{1}{2\pi Gg\textrm{cosh}2s}e^{-\frac{1}{2\textrm{cosh}2s}(x_3^2+y_3^2)}
\cr &&\hskip 0.3 in\times W_{in}(\frac{g}{G}x_3\textrm{tanh}2s+\frac{i_{X'_1}}{G}, \frac{G}{g}y_3\textrm{tanh}2s-\frac{i_{Y'_2}}{g}).~~~~
\end{eqnarray}
When $g\gg1$ so that $G=\sqrt{1+g^2}\approx g$, and $\cosh 2s \gg$ the range of $W_{i}(x_3,y_3)$, we have
\begin{eqnarray}
&& W_{proj}(x_3,y_3) \cr &&\hskip 0.4in \simeq  \frac{1}{2\pi Gg\textrm{cosh}2s}W_{in}(x_3+\frac{i_{X'_1}}{G},y_3-\frac{i_{Y'_2}}{g}).~~~~
\end{eqnarray}
With the detection outcomes of $i_{X'_1},i_{Y'_2}$, we transmit these measurement result  through a classical channel to the location of field 3 and a displacement operation of $x_3+\frac{i_{X'_1}}{G} \rightarrow x_3, y_3-\frac{i_{Y'_2}}{g} \rightarrow y_3$ (${\rm M}_X, {\rm M}_Y$ in Fig.\ref{fig:ill}) can be performed on field 3, leading to the displaced Wigner function as
\begin{eqnarray}
 W_{proj}^{disp}(x_3,y_3) \propto   W_{in}(x_3,y_3).
 \end{eqnarray}
This recovers the Wigner function of the input state thus achieving quantum state teleportation. So, we just showed that the quantum state teleportation scheme still works even after we replace the BS with a high gain PA. Note that the condition $G=\sqrt{1+g^2}\approx g$ is equivalent to $t=r=1/\sqrt{2}$, which is required for the scheme with a BS.

\section{Tolerance to detection loss}

The quantum teleportation process involves homodyne detection which may introduce losses through detectors' less-than-unit quantum efficiency and imperfect mode matching to the local oscillator fields. It was known that PA-assisted homodyne measurement is tolerant to detection and propagation losses \cite{jml19}. We will discuss the influence of these losses in this section. Let us start with the traditional BS scheme. This case was treated in Ref.\cite{brau2} but we present it here again for the sake of comparison with the PA case. For the input state in Eq.(\ref{W-in}) and from Eq.(\ref{BS:Win-out}), the Wigner function
after the 50:50 beam splitter of Bell measurement is
\begin{widetext}
\begin{eqnarray}
&&W_{out}^{(BS)}(x'_1,y'_1;x'_2,y'_2;x_3,y_3) \cr &&\hskip 0.2in =\frac{1}{(2\pi)^2}e^{-\frac{1}{4}[(x_3+\frac{x'_1+x'_2}{\sqrt{2}})^2+
(y_3-\frac{y'_1+y'_2}{\sqrt{2}})^2]e^{2s}}
e^{-\frac{1}{4}[(x_3-\frac{x'_1+x'_2}{\sqrt{2}})^2+(y_3+\frac{y'_1+y'_2}{\sqrt{2}})^2]e^{-2s}}
W_{in}\left(\frac{x'_1-x'_2}{\sqrt{2}},\frac{y'_1-y'_2}{\sqrt{2}}\right).
\end{eqnarray}
\end{widetext}
We now introduce detection losses by using a beam splitter model with the same transmissivity $\eta$ for both output field 1 and 2 right before detection. Since the homodyne detection is on the $x$-quadrature of field 1 and $y$-quadrature of field 2, we leave $y_1',x_2'$ unchanged and only consider effect on $x'_1,y'_2$ together with the vacuum from the unused port of the beam splitter:
$x'_1\xrightarrow{}\eta x''_1+\sqrt{1-\eta^2}x'_v$, $y'_2\xrightarrow{}\eta y''_2+\sqrt{1-\eta^2}y'_v$,
$x_v\xrightarrow{}\eta x'_v-\sqrt{1-\eta^2}x_1''$,
$y_v\xrightarrow{}\eta y'_v-\sqrt{1-\eta^2}y_2''$. Here, $\hat{a}_v'$ is the vacuum coupled in through loss. Then after including vacuum Wigner function, the Wigner function before the homodyne detection is
\begin{widetext}
\begin{eqnarray}
&&W_{out}^{(BS)}(x''_1,y'_1;x'_2,y''_2;x_3,y_3)\cr &&\hskip 0.2in =\frac{1}{(2\pi)^3}e^{-\frac{1}{4}[(x_3+\frac{\eta x''_1+\sqrt{1-\eta^2}x'_v+x'_2}{\sqrt{2}})^2+(y_3-\frac{y'_1+\eta y''_2+\sqrt{1-\eta^2}y'_v}{\sqrt{2}})^2]e^{2s}} e^{-\frac{1}{4}[(x_3-\frac{\eta x''_1+\sqrt{1-\eta^2}x'_v+x'_2}{\sqrt{2}})^2+(y_3+\frac{y'_1+\eta y''_2+\sqrt{1-\eta^2}y'_v}{\sqrt{2}})^2]e^{-2s}}
\cr &&\hskip 0.5in \times W_{in}\left(\frac{\eta x''_1+\sqrt{1-\eta^2}x'_v-x'_2}{\sqrt{2}},\frac{y'_1-\eta y''_2-\sqrt{1-\eta^2}y'_v}{\sqrt{2}}\right) e^{-\frac{1}{2}[(\eta x'_v-\sqrt{1-\eta^2}x_1'')^2+(\eta y'_v-\sqrt{1-\eta^2}y_2'')^2]}.
\end{eqnarray}
\end{widetext}

For the homodyne measurement with result of  $x''_1=i_{X'_1}$, $y''_2=i_{Y'_2}$, we obtain the Wigner function for the projected state of field 3 by setting $x''_1=i_{X'_1}$, $y''_2=i_{Y'_2}$ and  integrating the variables $y'_1, x'_2, x_v',y_v'$. In the limit of large $s$, the projected Wigner function is
\begin{eqnarray}\label{Wproj-BS}
&&W_{proj}^{(BS)}(x_3,y_3)\cr &&\hskip 0.2in
\propto \int \textrm{d}x\textrm{d}y W_{in}(x,y) e^{-\frac{1}{2{\sigma}_1^2}[(x-\frac{\sqrt{2}i_{X'_1}}{\eta}-x_3)^2
+(y-\frac{\sqrt{2}i_{Y'_2}}{\eta}-y_3)^2]}.\cr &&
\end{eqnarray}
where ${\sigma}_1^2=2(\frac{1-\eta^2}{\eta^2}+e^{-2s})$. In the ideal case of no loss, we have  $\eta=1$ and for large $s$, the Gaussian in Eq.(\ref{Wproj-BS}) becomes a $\delta$-function so that after the required displacement of $x_3+\frac{\sqrt{2}i_{X'_1}}{\eta} \rightarrow x_3, y_3+\frac{\sqrt{2}i_{Y'_2}}{\eta}\rightarrow y_3$ upon receiving of the detection outputs $i_{X'_1},i_{Y'_2}$, we recover the input Wigner function $W_{i}(x_3,y_3)$. But with finite detection losses, vacuum noise will come into the quantum teleportation channel and even in the limit of large $s$, we have
\begin{eqnarray}\label{Wproj-BS2}
&&W_{proj}^{(BS)}(x_3,y_3)\cr &&\hskip 0.2in
\propto \int \textrm{d}x\textrm{d}y W_{in}(x,y) e^{-\frac{\eta^2}{4(1-\eta^2)}[(x-x_3)^2
+(y-y_3)^2]},~~~~
\end{eqnarray}
which involves a convolution with the vacuum Wigner function. This will reduce the fidelity of teleportation (see next section). The expression in Eq.(\ref{Wproj-BS2}) was first derived in Ref.\cite{brau2}.

For the scheme with a PA in place of the BS, we introduce losses after the output of the PA but before the homodyne measurement. From Eq.(\ref{W-PA}) and similar to Eq.(\ref{Wproj-BS}), we have the projected Wigner function of field 3 after the homodyne measurement of results $x_1'=i_{X'_1}, y_2'=i_{Y'_2}$:
\begin{widetext}
\begin{equation}
\begin{split}
W_{proj}^{(PA)}(x_3,y_3)&=\frac{1}{(2\pi)^3}e^{-\frac{1}{2\textrm{cosh}2s}(x_3^2+y_3^2)}\int dx_2'dy_1'dx'_vdy'_ve^{-\frac{\textrm{cosh}2s}{2}(Gx'_2+x_3\textrm{tanh}2s-g(\eta i_{X'_1}+\sqrt{1-\eta^2}x'_v))^{2}}\\&\hskip 0.2in\times e^{-\frac{\textrm{cosh}2s}{2}(gy'_1+G(\eta i_{Y'_2}+\sqrt{1-\eta^2}y'_v)-y_3\textrm{tanh}2s)^2}e^{-\frac{1}{2}[(\eta x'_v-\sqrt{1-\eta^2}i_{X'_1})^2+(\eta y'_v-\sqrt{1-\eta^2}i_{Y'_2})^2]}\\&\hskip 0.4in\times W_{in}(G(\eta i_{X'_1}+\sqrt{1-\eta^2}x'_v)-gx'_2,Gy'_1+g(\eta i_{Y'_2}+\sqrt{1-\eta^2}y'_v))
\end{split}
\end{equation}
Integrating $y'_1$, $x'_2$ via $\delta$-function approximation in the large $s$ limit, we obtain the projected Wigner function
\begin{eqnarray}\label{W-proj-PA}
&&W_{proj}^{(PA)}(x_3,y_3)
\propto \int \textrm{d}x\textrm{d}y W_{in}(x,y) e^{-\frac{1}{2{\sigma}_2^2}[(x-\frac{i_{X'_1}}{\eta G}-\frac{gx_3}{G})^2+(y+\frac{i_{Y'_2}}{\eta g}-\frac{Gy_3}{g})^2]}.
\end{eqnarray}
\end{widetext}
where
${\sigma}_2^2=2(\frac{1-\eta^2}{\eta^2}\times\frac{1}{2G^2}+e^{-2s}\times\frac{g^2}{G^2})$. Setting $G\gg 1$ so that $G\approx g$ and making the displacement operation of $x_3 +i_{X'_1}/\eta g \rightarrow x_3$ and $y_3 - i_{Y'_2}/\eta G \rightarrow y_3 $, we have
\begin{eqnarray}
&&W_{proj}^{(PA)}(x_3,y_3)\cr &&\hskip 0.3in \propto \int \textrm{d}x\textrm{d}y W_{in}(x,y) e^{-\frac{1}{2{\sigma}_2'^2}[(x-x_3)^2+(y-y_3)^2]},~~~~
\end{eqnarray}
where ${\sigma}_2'^2\approx 2(\frac{1-\eta^2}{2\eta^2G^2}+e^{-2s})$. If $G^2 \gg  (1-\eta^2)e^{2s}/2\eta^2$, we have ${\sigma}_2'^2\rightarrow 2e^{-2s}$, which is the lossless case ($\eta = 1$) of Eq.(\ref{Wproj-BS}). Therefore, the effect of losses can be mitigated by large $G$.

\section{Influence of losses via Fidelity}

To quantify the influence of losses  and the gain size of PA in place of BS, we consider the quantity of fidelity. We will calculate the entanglement fidelity \cite{ent-fi}, which quantifies how well a quantum (teleportation) channel, which may interact with the environment E, preserves the transferred input state (in a state space denoted as Q) and its entanglement with another system in the space of R. The input state ($\hat \rho_{in}$) can be obtained by taking partial trace of an entangled pure state on a larger Hilbert space (a joint space of Q and the entangled space R) over the space R. For the case of no entanglement, the input state is then a pure state.   According to Ref.\cite{ent-fi}, the entanglement fidelity only depends on the initial quantum state and the dynamic evolution of the input state on Q through the quantum channel.  Suppose the general quantum evolution of the state on Q through the quantum channel can be cast in the form of $\hat{\rho}_{out}=\sum_k \hat{A}_k\hat{\rho}_{in}\hat{A}_k^\dagger$ by some operator-sum representation with $\hat{A}_i$ being a collection of operators acting in the space of Q and satisfying the completeness relation $\sum_k \hat{A}_k^\dagger\hat{A}_k=1$. The entanglement fidelity is defined as \cite{ent-fi}
\begin{equation}\label{ent-fid}
\begin{split}
F_e=\sum_k \textrm{Tr}(\hat{A}_k\hat{\rho}_{in})\textrm{Tr}(\hat{A}_k^\dagger\hat{\rho}_{in}).
\end{split}
\end{equation}
For the special case of a pure input state $\hat \rho_{in} = |\phi_{in}\rangle\langle \phi_{in}|$, we have $F_e = \sum_k |\langle \phi_{in}|\hat A_k|\phi_{in}\rangle|^2 = \langle \phi_{in}|\hat{\rho}_{out} |\phi_{in}\rangle$.

\subsection{Scheme with beam splitter}

Consider the quantum state teleportation scheme as the quantum channel. For the case of using a BS for Bell measurement and with large $s$, the Wigner functions of the output is connected to the input by Eq.(\ref{Wproj-BS2}) and rewritten as
\begin{equation}\label{Wout-in}
\begin{split}
W_{out}= W_{in}\circ G_{{\sigma}}
\end{split}
\end{equation}
with $\circ$ denoting convolution and $G_{{\sigma}}$ as the two dimensional Gaussian distribution with variance ${\sigma}^2 = 2(1-\eta^2)/\eta^2$. The teleportation input-output relation for the Wigner functions described by Eq.(\ref{Wout-in}) can also be cast in the density operator form as (See Appendix)
\begin{equation}\label{den-inout}
\begin{split}
\hat{\rho}_{out}= \int dx dy \hat{D}\Big(\frac{x+yj}{2}\Big)\hat{\rho}_{in}\hat{D}^{\dagger}\Big(\frac{x+yj}{2}\Big) G_{{\sigma}}(x,y),
\end{split}
\end{equation}
where operator $\hat{D}(\alpha) \equiv \exp(\alpha \hat a^{\dag}-\alpha^*\hat a)$ is the displacement operator ($\alpha = (x+jy)/2$).
The expression in Eq.(\ref{den-inout}) is in the operator-sum representation form required for Eq.(\ref{ent-fid}). Here,
since the density operator is represented on the infinite dimensional Hilbert space, the summation sign is replaced by a two dimensional integral over $x, y$ and the evolution operator $\hat{A}_i$ is replaced by $\hat{A}(x,y)=\hat{D}(\frac{x+yj}{2})\sqrt{G_{{\sigma}}(x,y)}$  in Eq.(\ref{den-inout}), which satisfies the completeness relation $\int dx dy \hat{A}^\dagger(x,y)\hat{A}(x,y)=1$. Therefore, the entanglement fidelity defined in Eq.(\ref{ent-fid}) is changed to
\begin{eqnarray}\label{fid-cal}
F_e&=&\int dx dy \textrm{Tr}[\hat{A}(x,y)\hat{\rho}_{in}]\textrm{Tr}[\hat{A}^\dagger(x,y)\hat{\rho}_{in}]\cr&=& \int dx dy |\chi_{in}(x,y)|^2 G_{{\sigma}}(x,y)
\end{eqnarray}
where $\chi_{in}(x,y)\equiv\textrm{Tr}  [\hat{D}(\frac{x+yj}{2})\hat{\rho}_{in}]$ is the characteristic function defined in Eq.(\ref{chi})
for the input state's Wigner function.

\begin{figure}
\begin{center}
\includegraphics[width=3.0in]{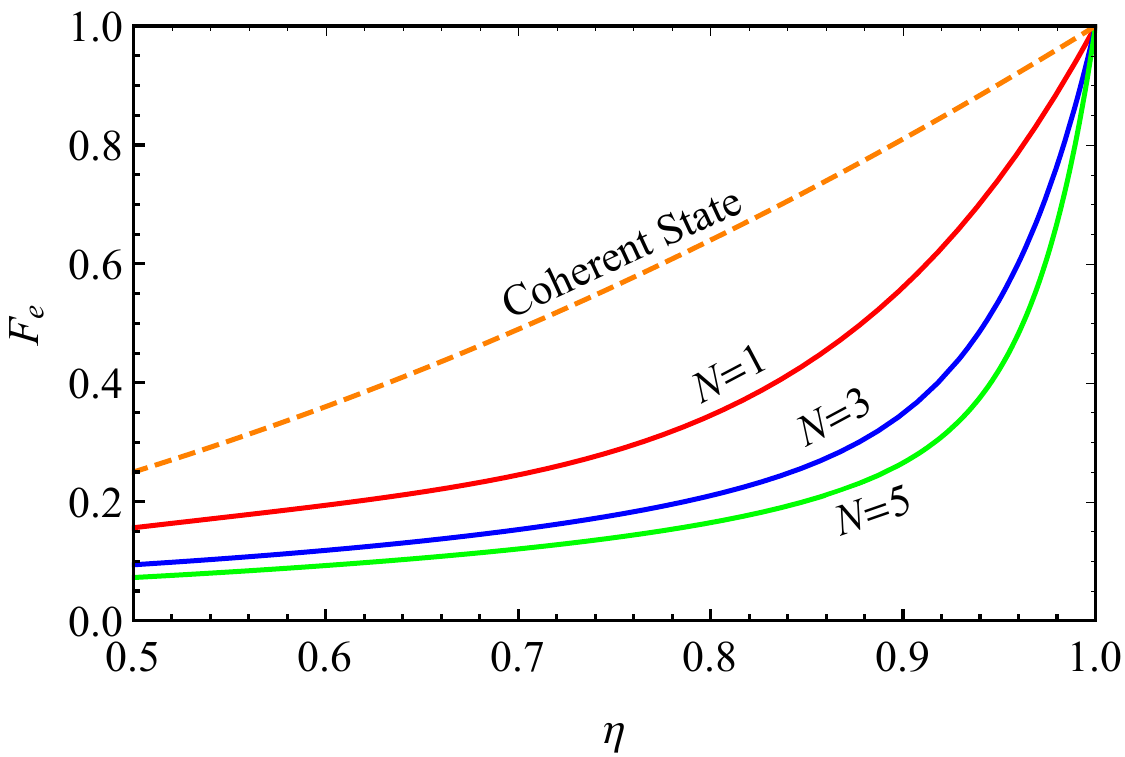}
\end{center}
\caption{Entanglement fidelity as a function of detection loss in the teleportation scheme with a BS for a coherent state $|\alpha\rangle$ with $\alpha = 3+3j$ (dashed line) and Fock states of $N=1,3,5$.}
\label{Fig-coh}
\end{figure}

We can now evaluate the entanglement fidelity for a number of known input states. First for coherent state, it is easy to obtain the fidelity as
\begin{equation}\label{fid-coh-BS}
F_e = \frac{1}{1+{\sigma}^2/2}
\end{equation}
with ${\sigma}^2 = 2(1-\eta^2)/\eta^2$. For Fock state $|N\rangle$,
\begin{eqnarray}
&&|\chi_{in}(x,y)|^2=|\langle N|\hat{D}\Big(\frac{x+yj}{2}\Big)|N\rangle |^2\cr &&\hskip 0.2in =\left |\int \textrm{d}x' f^*_N(x')f_N(x'-x/\sqrt{2})e^{j\frac{y(x'-x/\sqrt{2})}{\sqrt{2}}}\right |^2,~~~~
\end{eqnarray}
where
\begin{equation}\label{f-N}
\begin{split}
 f_N(x)=\frac{\pi^{-1/4}}{\sqrt{2^N N!}}e^{-x^2/2}H_N(x)
\end{split}
\end{equation}
with $H_N(x)$ as the N$th$-order Hermite polynomials.
We can then evaluate the entanglement fidelity $F_e$ with Eq.(\ref{fid-cal}) for a given loss modeled by a BS with transmission coefficient $\eta$. Figure \ref{Fig-coh} plots the dependence of $F_e$ as a function of $\eta$ for a coherent state of $\alpha=3+3j$ and Fock states of $N=1,3,5$, showing the fast drop of $F_e$ with the increase of loss (decrease of $\eta$). The rate of drop is especially large for number states with higher photon numbers as compared to the coherent state (dashed curve). Thus, nonclassical states are more sensitive to loss in the teleportation process.

\subsection{Scheme with parametric amplifier}

Next we consider the teleportation scheme with aid of a PA but having losses $1-\eta$ before detection. The output is related to input by Eq.(\ref{W-proj-PA}) and can be rewritten in the form of Eq.(\ref{Wout-in}) as
\begin{equation} \label{W-out-PA}
\begin{split}
W_{out}= \int \textrm{d}x\textrm{d}y W_{in}(x\frac{g}{G}-x',y\frac{G}{g}-y') G_{\Bar{\sigma}}(x',y'),
\end{split}
\end{equation}
where $\bar\sigma^2\equiv (1-\eta^2)/\eta^2G^2$. Then Eq.(\ref{den-inout}) is changed to (See Appendix)
\begin{eqnarray}\label{rho-out-PA}
\hat{\rho}_{out} &&= \int dx dy \hat{S}(\epsilon)\hat{D}(\frac{x+yj}{2})\hat{\rho}_{in}\cr
&&\hskip 0.8in \times\hat{D}^{\dagger}(\frac{x+yj}{2})\hat{S}^{\dagger}(\epsilon) G_{\Bar{\sigma}}(x,y),
\end{eqnarray}
where $\hat{S}(\epsilon)\equiv \exp[\epsilon(\hat a^{\dag 2}-\hat a^2)/2]$ is the squeezing operator with $\epsilon \equiv \mathrm{ln}(G/g)$, and Eq.(\ref{fid-cal}) is modified to
\begin{equation}\label{fid-cal-PA}
\begin{split}
F_e= \int dx dy |\chi_{PA}(x,y)|^2 G_{\Bar{\sigma}}(x,y)
\end{split}
\end{equation}
with $\chi_{PA}(x,y)=\textrm{Tr}[\hat{S}(\epsilon)  \hat{D}(\frac{x+yj}{2})\hat{\rho}_{in}]$.

\begin{figure}
\begin{center}
\includegraphics[width=3.1in]{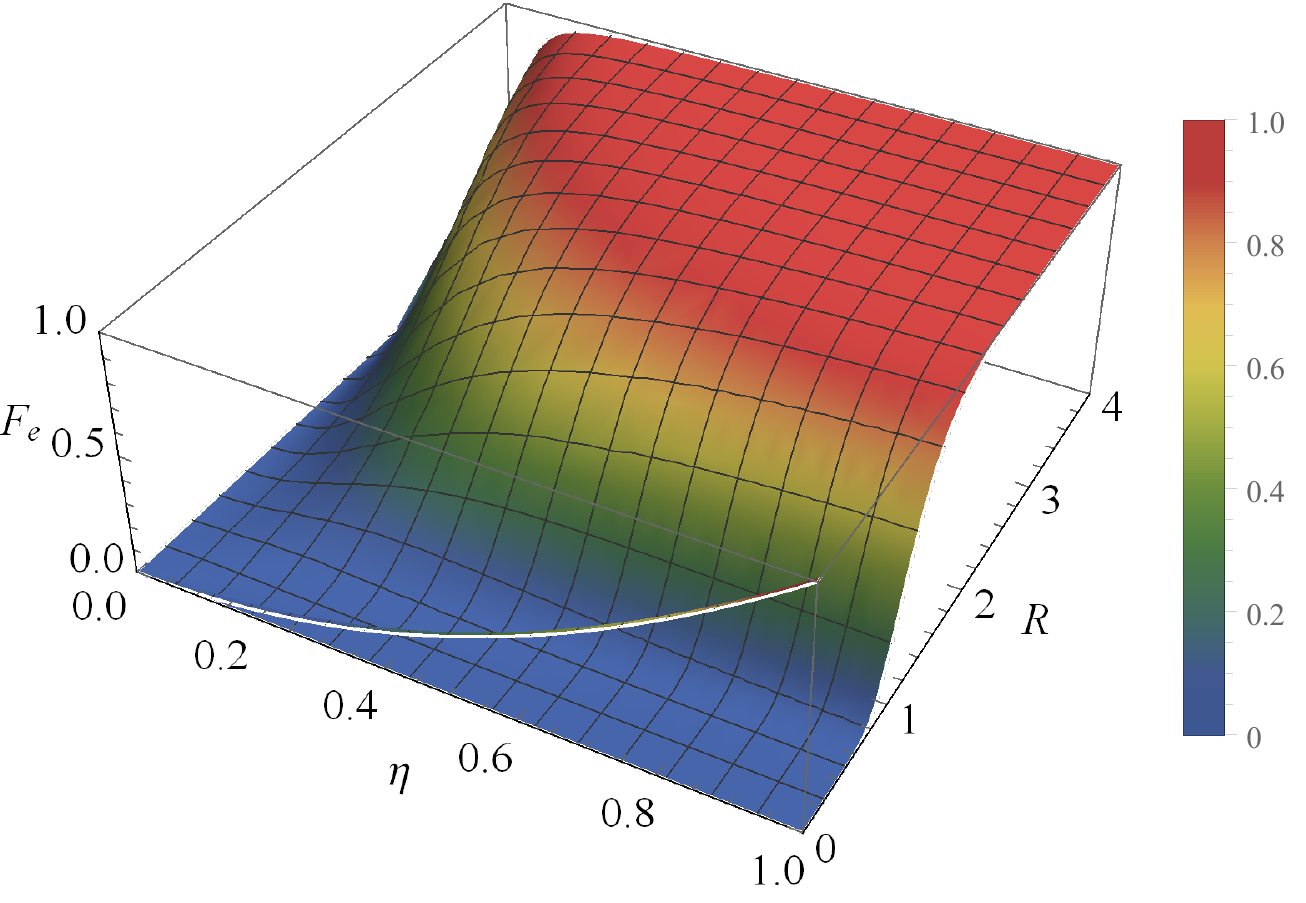}
\end{center}
\caption{Three-dimensional plot of entanglement fidelity $F_e$ as a function of transmission $\eta$ (opposite of loss) and gain parameter $R (G\equiv \cosh R)$ of the parametric amplifier used in Bell measurement for teleportation of the coherent state $|\alpha\rangle$ with $\alpha = 3+3j$. The light curve at $R=0$ corresponds to the case of using a beam splitter for the Bell measurement.}
\label{Fig-coh1}
\end{figure}

For a coherent state $|\alpha\rangle$, we have
\begin{eqnarray}
|\chi_{PA}^{(\alpha)}(x,y)|^2&=&|\langle \alpha|\hat{S}(\epsilon)\hat{D}(\frac{x+yj}{2})
|\alpha\rangle|^2\cr &=&|\bra{\alpha}\hat{S}(\epsilon)\ket{\alpha+\frac{x+yj}{2}}|^2.
\end{eqnarray}
Setting ${\alpha}=a+bj$, we then obtain from Eq.(\ref{fid-cal-PA}) with some manipulation
\begin{eqnarray}\label{fid-coh-PA}
F_e&=&\frac{\exp\left[\frac{a^2(\frac{\mu+\nu-1}{\mu})^2}{1+
\frac{2}{\bar{\sigma}^2}+\frac{\nu}{\mu}}+
\frac{b^2(\frac{\mu-\nu-1}{\mu})^2}{1+\frac{2}{\bar{\sigma}^2}-
\frac{\nu}{\mu}}-2\frac{\mu-1}{\mu}(a^2+b^2)\right]}
{\mu\bar{\sigma}^2\sqrt{(\frac{1}{2}
+\frac{1}{\bar{\sigma}^2})^2-
(\frac{\nu}{2\mu})^2}}\cr
&\approx & \frac{\exp\left[-\frac{a^2+b^2}{1+\bar\sigma^2/2}(\frac{\nu}{\mu})^2\right]}
{1+\bar\sigma^2/2} ~~{\rm for} ~~ G\gg 1,
\end{eqnarray}
where $\mu\equiv \textrm{cosh}(\epsilon) = (G^2+g^2)/2Gg, \nu \equiv \textrm{sinh}(\epsilon)= 1/2Gg$. When $G$ tends to a large value, we have $G\sim g$ and $\epsilon \sim 0, \mu\sim 1, \nu\sim 0$ and Eq.(\ref{fid-coh-PA}) approaches Eq.(\ref{fid-coh-BS}) but with $\sigma^2$ replaced by $\bar \sigma^2 \equiv 2\frac{1-\eta^2}{\eta^2}\times \frac{1}{2G^2}$, which goes to zero as $G$ becomes large. Hence, $F_e\rightarrow 1$ for large $G$ and independent of the loss $\eta$. So, with the aid of a PA of large gain, the effect of detection loss can be reduced to zero. This is demonstrated in Fig.\ref{Fig-coh1} as the red region ($F_e\sim 1$) in the 3-D plot of $F_e$ as a function of $\eta$ and the gain-related parameter $R$ with $G\equiv \cosh R$ (or $R\equiv \ln(G+\sqrt{G^2-1})$). Figure \ref{Fig-coh1} is obtained from the first expression in Eq.(\ref{fid-coh-PA}) without approximation. The red region extends to low value of $\eta$ ($<0.5$, large loss) at high gain ($R>2$). The light colored curve at $R=0$ is for the case when we use a beam splitter for Bell measurement. As can be seen, $F_e$ drops fast as $\eta$ decreases.

\begin{figure}
\begin{center}
\includegraphics[width=3.1in]{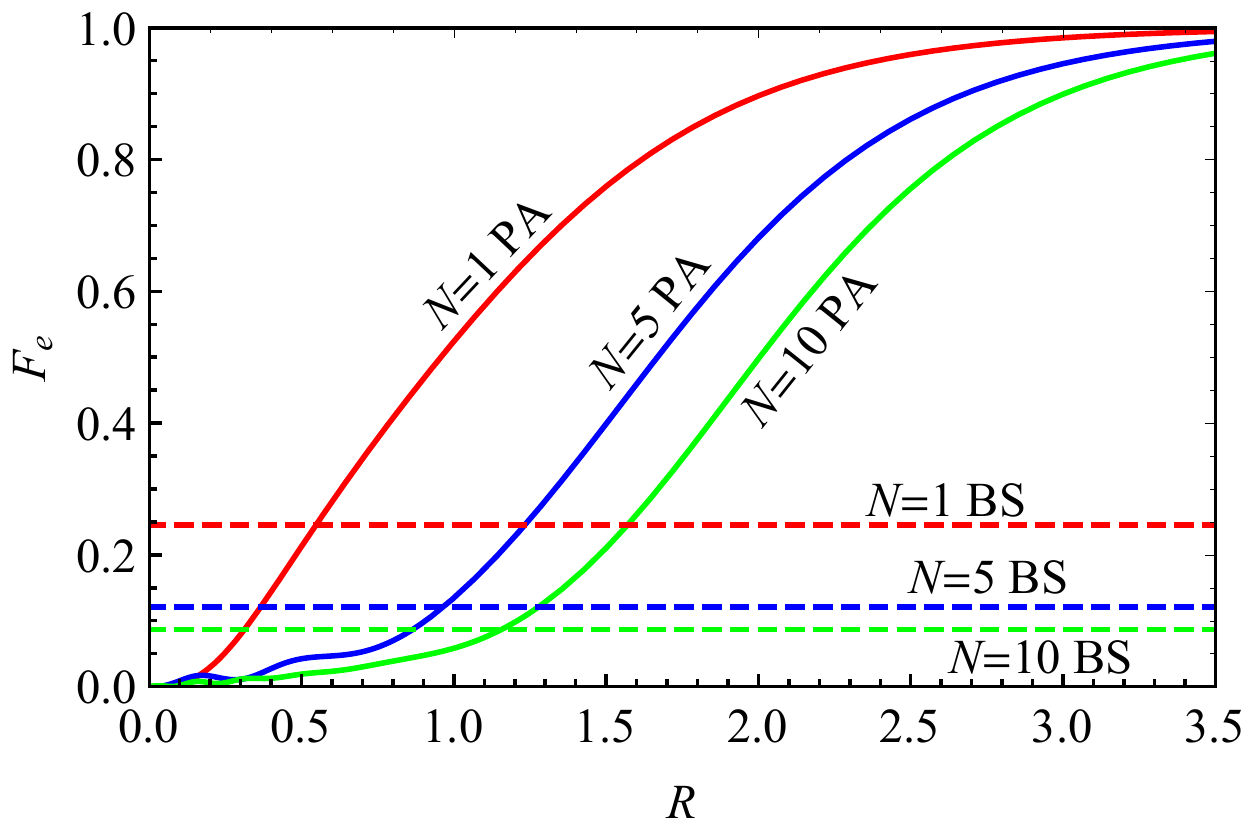}
\end{center}
\caption{Entanglement fidelity as a function of gain of parametric amplifier for at a loss of $\eta=0.7$ for Fock state $|N\rangle$ with $N=1,5,10$. The dashed lines are for the BS scheme.}
\label{Fig-FS}
\end{figure}

On the other hand, even with no detection loss ($\eta=1, \bar\sigma=0$) but a finite $G$, we have from Eq.(\ref{fid-coh-PA})
\begin{eqnarray}\label{fid-coh-PA2}
F_e &=&\frac{1}{\mu}\exp\left[-2\frac{\mu-1}{\mu}(a^2+b^2)\right] \cr
& \approx &\exp\left[-(a^2+b^2)/4G^4\right] ~~{\rm for} ~~ G\gg 1
\end{eqnarray}
The blue region (low $F_e<0.15$) in Fig.\ref{Fig-coh1} extends to high $\eta$ value when $R<1$ for relatively low gain, which indicates that high gain ($R>2$) is required for the PA-assisted scheme. From Eq.(\ref{fid-coh-PA2}), we find that in order to have $F_e\approx 1$, we need $G^2 \gg \sqrt{a^2+b^2} = |\alpha|$, that is, the larger the average the photon number, the bigger the gain $G$ needs to be. This behavior is not limited to coherent states as we will see next for photon number Fock states.

\begin{figure}
\begin{center}
\includegraphics[width=3.1in]{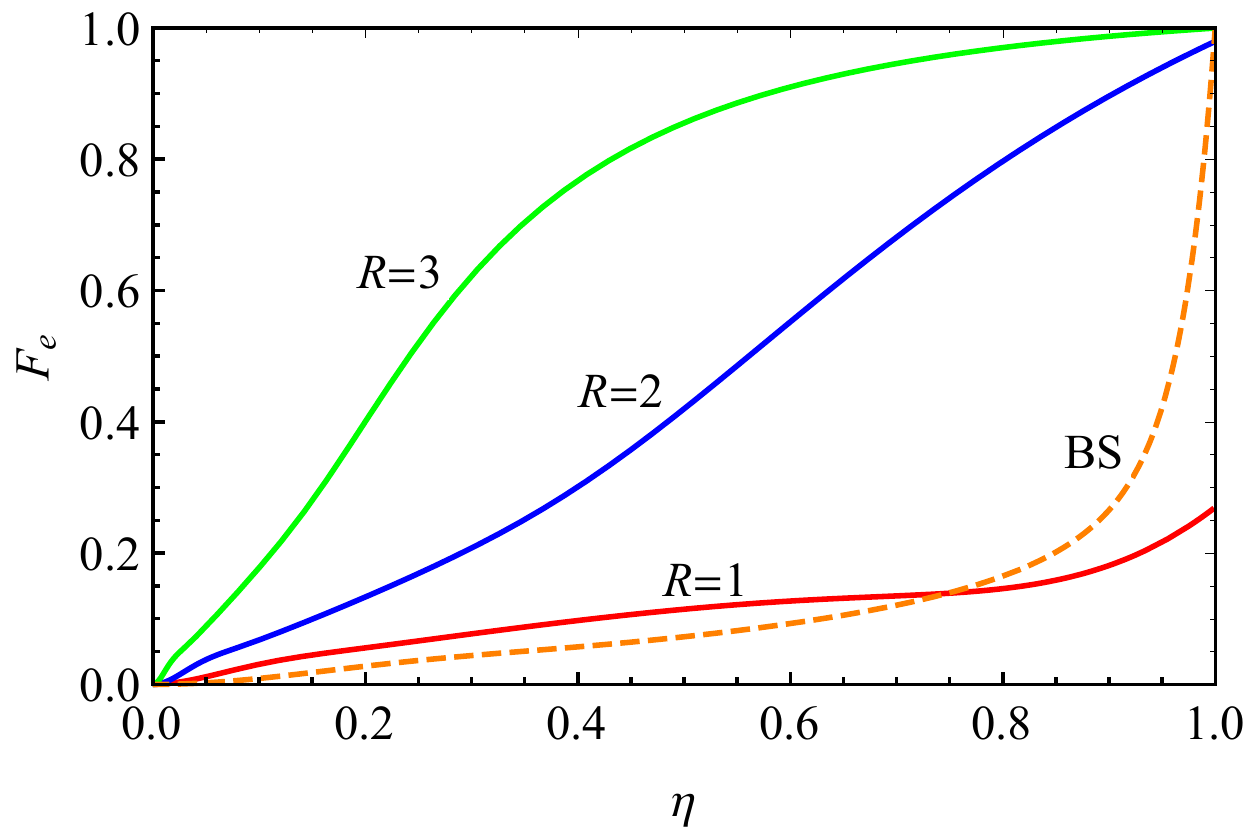}
\end{center}
\caption{Entanglement fidelity $F_e$ as a function of transmission $\eta$ for Fock state $|N\rangle$ with $N=5$ at gain of $R=1,2, 3$ for the PA-assisted scheme. The case of BS scheme is plotted as the dashed line for comparison.}
\label{Fig-FS2}
\end{figure}

Next we look at the nonclassical states of Fock state $\ket{N}$. The characteristic function $\chi_{PA}$ in Eq.(\ref{fid-cal-PA}) has the form of
\begin{eqnarray}
|\chi_{PA}^{(N)}(x,y)|^2&=&{|\bra{N}\hat{S}(\epsilon)\hat{D}(\frac{x+yj}{2})\ket{N}|^2}\cr&=& {\Big |\int \textrm{d}x'\sqrt{\mathrm{tanh}R}~e^{-jy[(x'/\sqrt{2})\mathrm{tanh}
(\epsilon)+x/2]}}\cr&&\hskip 0.05in {\times  f^*_N(x')f_N(-x'\mathrm{tanh}(\epsilon)-x/\sqrt{2}) \Big |^2,}~~~~~~
\end{eqnarray}
where the definition of $f_N(x)$ is the same as Eq.(\ref{f-N}). The fidelity can be calculated numerically from Eq.(\ref{fid-cal-PA}). We plot in Fig.\ref{Fig-FS} the fidelity $F_e$ as a function of the gain-related parameter $R$ for Fock states $|N\rangle$ with  $N=1,5,10$, respectively. The detection loss is set with transmission $\eta =0.7$. As can be seen, larger gain ($R$ value) is needed for higher $N$ to reach $F_e\approx 1$, similar to the case of coherent states as predicted by the second line of Eq.(\ref{fid-coh-PA2}). We also plot in Fig.\ref{Fig-FS} the corresponding values of $F_e$ for the BS scheme (dashed lines) for comparison, demonstrating the effect of PA to counter the detrimental effect of detection loss. The effect of loss on the Fock state $|5\rangle$ is displayed in Fig.\ref{Fig-FS2}, where we plot $F_e$ as a function of transmission coefficient $\eta$ for three values of $R$. The result of the BS scheme (dashed line) is also plotted for comparison. As expected, PA-assisted scheme is no good for the case of relatively low gain ($R=1,2$). But with $R=3$, it keeps relatively high $F_e$ value ($>0.8$) even at a large loss of 50\% ($\eta=0.5$).

\section{Influence of loss on Entanglement}
The input states in the previous sections are all pure states. In quantum communication, we more often transmit entangled states. We will examine how losses in the two teleportation schemes will affect the transmission of an EPR-type of entangled states which is simply a two-mode squeezed state with a Wigner function given in Eq.(\ref{EPR}).

\subsection{Inseparability}
We first consider the inseparability quantity $I_s$ defined as \cite{duan00}
\begin{eqnarray}
I_s \equiv \langle\Delta^2(\hat X_1-\hat X_2)\rangle + \langle\Delta^2(\hat Y_1+\hat Y_2)\rangle.
\end{eqnarray}
For un-entangled fields, it has a minimum value of $I_s^{(0)}=4$ for vacuum. $I_s < I_s^{(0)}=4$ gives the criterion for entanglement between two fields and the smaller the value of $I_s$ is, more entangled are the two fields. The ideal value is $I_s=0$, showing perfect EPR correlation between $\hat X_1,\hat X_2$ and between $\hat Y_1,\hat Y_2$. For the EPR entangled state given in Eq.(\ref{EPR}) with $s=-1$, we have normalized value $I_s^{EPR}/I_s^{(0)} = 0.135 = - 8.69$dB. We will teleport one of the two entangled fields, say the signal beam,  through the BS or PA-assisted teleportation scheme.

\begin{figure}
\begin{center}
\includegraphics[width=3.1in]{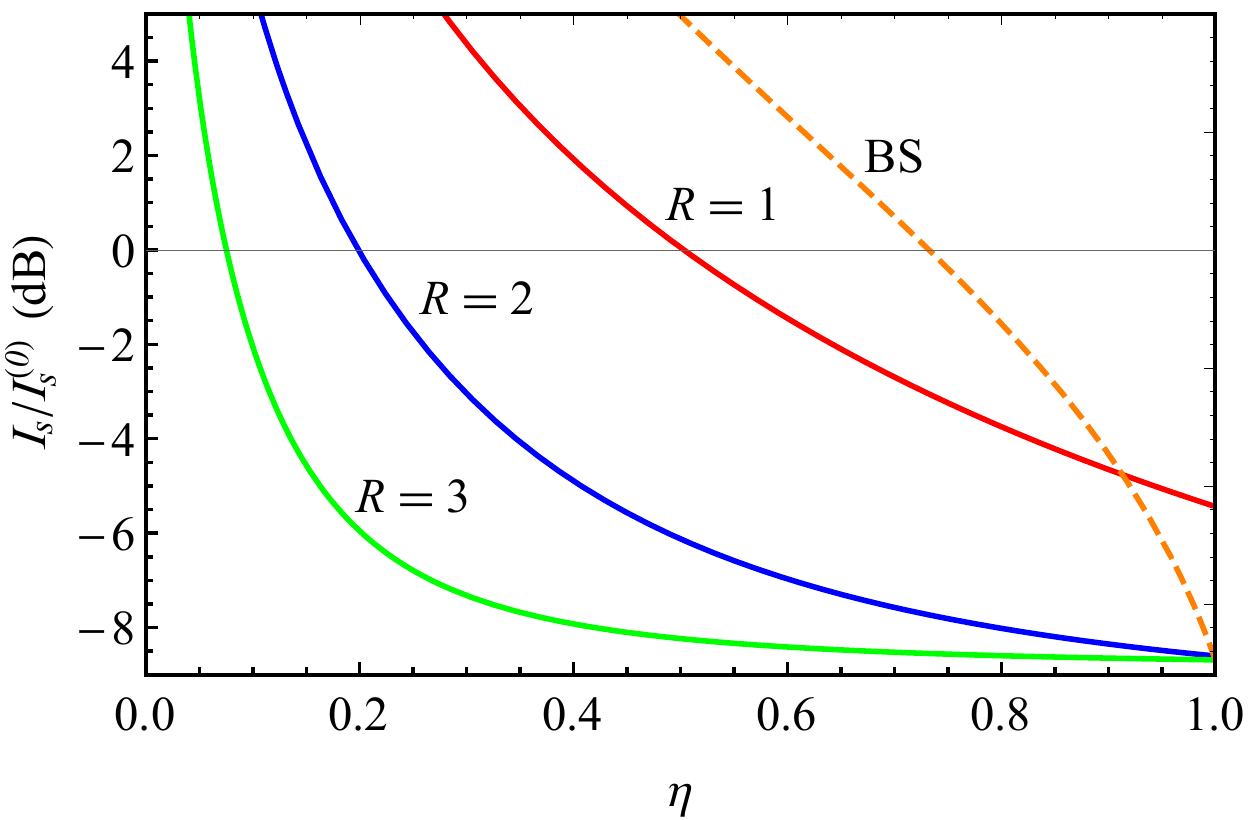}
\end{center}
\caption{Normalized inseparability $I_s/I_s^{(0)}$ ($I_s^{(0)}=2$ for vacuum) in log-scale as a function of transmission $\eta$ for the EPR-entangled state with initial input $I_s^{EPR}/I_s^{(0)} = 0.135 =- 8.69 $ dB for various gain parameters of $R=1,2, 3$ for the PA-assisted scheme (solid) and the BS scheme (dashed). }
\label{Fig-Is}
\end{figure}

\begin{figure}
\begin{center}
\includegraphics[width=3.1in]{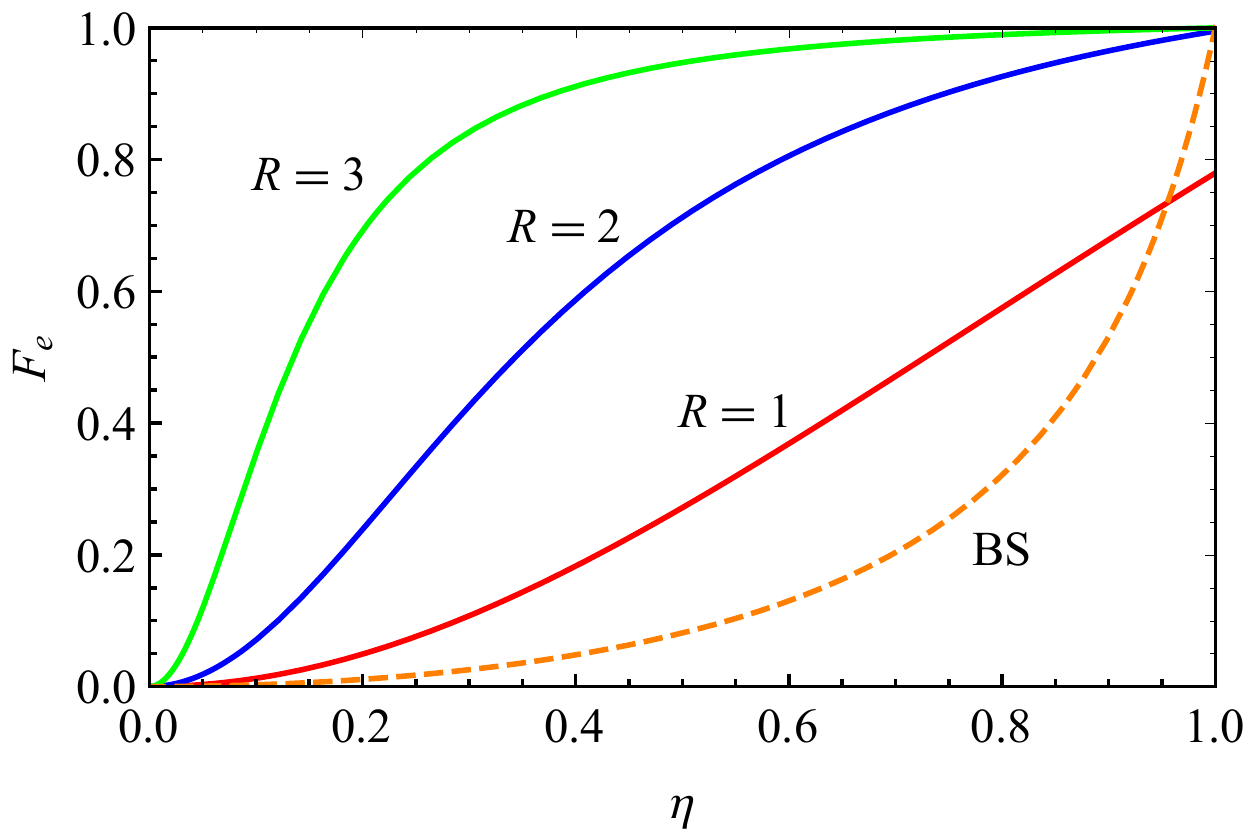}
\end{center}
\caption{Entanglement fidelity $F_e$ as a function of transmission $\eta$ for a thermal state with average photon number $\bar n =\sinh^2(-1)=1.38$ for the PA-assisted scheme with gain parameters of $R=1,2, 3$ (solid)  and for the BS scheme (dashed).}
\label{Fig-th}
\end{figure}

The Wigner functions of the output state are given by Eq.(\ref{Wout-in}) and (\ref{W-out-PA}) as
\begin{equation}\label{W-out-in2}
\begin{split}
W_{out}= \int \textrm{d}x'\textrm{d}y' W_{in}(x_1,y_1;x_2'-x',y_2'-y')G_{\sigma'}(x',y'),
\end{split}
\end{equation}
with $x_2'=x_2$, $y_2'=y_2$, $\sigma'=\sigma$ for the BS scheme and $x_2'=x_2\frac{g}{G}$, $y_2'=y_2\frac{G}{g}$, $\sigma'=\Bar{\sigma}$ for the PA-assisted scheme. We calculate  $I_s^{out}$ between the teleported signal field and the original idler field to examine how entanglement is affected by teleportation.
The inseparability quantity
$I_s^{out}=\braket{\Delta^2X_{-}}_{out}+\braket{\Delta^2Y_{+}}_{out}$ with $X_{-}=x_1-x_2$ and $Y_{+}=y_1+y_2$ is calculated from Wigner function by
\begin{eqnarray}
\braket{A}_{out} &=& \int \textrm{d}x_1\textrm{d}y_1\textrm{d}x_2\textrm{d}y_2 A(x_1,y_1;x_2,y_2)\cr
&&\hskip 0.5in \times W_{out}(x_1,y_1;x_2,y_2),
\end{eqnarray}
where $A=\Delta^2X_{-},\Delta^2Y_{+}$, respectively. $W_{out}$ is obtained from Eq.(\ref{W-out-in2}) with $W_{in}$ given in Eq.(\ref{EPR}) for an EPR entangled state. Figure \ref{Fig-Is} shows the results of calculation. As can be seen, the BS scheme (dashed curve) is very sensitive to losses: the value of $I_s$ increases quickly as detection efficiency $\eta$ drops and the fields are unentangled for $\eta<0.7$. The PA-assisted scheme, on the other hand, can keep $I_s$ at quite a low value  with a large gain ($R=3$) even for  $\eta$ as low as 0.5. Small gain cannot preserve the original $I_s$ value even at no loss $\eta =1$ but the fields are still entangled up to $\eta=0.5$.

\subsection{Fidelity}
The fidelity for entangled states can still be calculated as before like the pure states but we need to take partial trace of the idler component of the density operator of the entangled states. For the EPR state in Eq.(\ref{EPR}), the signal field becomes a thermal state with average photon number $\Bar{n}=\textrm{sinh}^2(s)$. Its density operator can be expressed with $P$-Representation as
$\hat{\rho}_{in}=\int \textrm{d}^2 \alpha P(\alpha)\ket{\alpha}\bra{\alpha}$, where $P(\alpha)=\frac{1}{\pi \Bar{n}}e^{-|\alpha|^2/\Bar{n}}$. The entanglement fidelity $F_e$ can be obtained from Eqs. (\ref{fid-cal}) and (\ref{fid-cal-PA}) with the characteristic functions being
\begin{eqnarray}
&&|\chi_{BS}(x,y)|^2\cr
&& \hskip 0.3in =\left|\int \textrm{d}^2 \alpha P(\alpha)\bra{\alpha}\hat{D}(\frac{x+yj}{2})\ket{\alpha}\right|^2\cr
&&\hskip 0.3in =\bigg|\int \textrm{d}^2 \alpha \frac{1}{\pi \Bar{n}}e^{-|\alpha|^2/\Bar{n}}\cr
 &&\hskip 0.7in \times e^{[(x+yj)\alpha^*-(x-yj)\alpha]/2-(x^2+y^2)/8}\bigg|^2, ~~~~
\end{eqnarray}
and
\begin{eqnarray}
&&|\chi_{PA}(x,y)|^2\cr
&&\hskip 0.3in=\left|\int \textrm{d}^2 \alpha P(\alpha)\bra{\alpha}\hat{S}(\epsilon)\hat{D}(\frac{x+yj}{2})\ket{\alpha}\right|^2\cr
&&\hskip 0.3in =\bigg|\int \textrm{d}^2 \alpha \frac{1}{\pi \Bar{n}}e^{-|\alpha|^2/\Bar{n}} e^{[(x+yj)\alpha^*-(x-yj)\alpha]/4}\cr
&&\hskip 1in \times \bra{\alpha}\hat{S}(\epsilon)\ket{\alpha+\frac{x+yj}{2}}\bigg|^2,
\end{eqnarray}
for the BS scheme and the PA-assisted scheme, respectively. Figure \ref{Fig-th} shows the results of calculation. It is very similar to Fig.\ref{Fig-FS2} for the number state case.

\section{Summary and Discussion}

In summary, we studied the quantum state teleportation scheme with a parametric amplifier (PA) replacing the beam splitter (BS) used in Bell measurement process. With large enough gain for the PA, the new scheme is as good as the original scheme. On the other hand, the employment of the PA can overcome the detection loss in the Bell measurement process, leading to a high teleportation fidelity even for a large detection loss. However, internal losses of PA and the losses before PA such as mode mis-match will be the losses imposed on the incoming fields and thus cannot be overcome by the employment of PA \cite{ou12}. They will have the same effect as in the BS scheme.

\begin{acknowledgments}
This work was supported  by US National Science Foundation (Grant No. 1806425).
\end{acknowledgments}




\vskip 0.3in

\appendix*
{\center{\bf APPENDIX}\\}

{\center{\bf Derivation of Eqs.(\ref{den-inout}) and (\ref{rho-out-PA})}}

\vskip 0.2in

For the scheme with a beam splitter for Bell measurement, we have from Eq.(\ref{Wout-in})
\begin{equation}
\begin{split}
W_{out}(X,Y)=\int W_{in}(X-x,Y-y)G_{\sigma}(x,y)\textrm{d}x\textrm{d}y.
\end{split}
\end{equation}
In terms of Wigner function, the density matrix is
\begin{eqnarray}
&&\hat{\rho}_{out}(\hat{X},\hat{Y})
\cr &&~~=\frac{1}{\pi}\int W_{out}(X,Y)e^{jv(\hat{X}-X)+ju(\hat{Y}-Y)}\textrm{d}v\textrm{d}u\textrm{d}X\textrm{d}Y
\cr &&~~=\frac{1}{\pi}\int W_{in}(X-x,Y-y)G_{\sigma}(x,y)
\cr && \hskip 0.4in \times e^{jv(\hat{X}-X)+ju(\hat{Y}-Y)}
\textrm{d}v\textrm{d}u\textrm{d}X\textrm{d}Y\textrm{d}x\textrm{d}y.~~~~~~
\end{eqnarray}
Now, let us shift $X, Y$ in $e^{jv(\hat{X}-X)+ju(\hat{Y}-Y)}$ to $X-x$, $Y-y$ by using operator $\hat D(\alpha) \equiv \exp(\alpha \hat a^{\dag}-\alpha^*\hat a)$: $\hat D(\alpha)\hat a\hat D^{\dag}(\alpha) = \hat a -\alpha$. With $\alpha=(x+jy)/2$,  we have
\begin{eqnarray}
&&\hat{\rho}_{out}(\hat{X},\hat{Y})
\cr &&~~~~=\frac{1}{\pi}\int W_{in}(X-x,Y-y)G_{\sigma}(x,y)
\cr &&\hskip 0.3in\times \hat{D}(\frac{x+yj}{2})e^{jv[\hat{X}-(X-x)]+ju[\hat{Y}-(Y-y)]}
\cr &&\hskip 0.5in\times \hat{D}^{\dagger}(\frac{x+yj}{2})\textrm{d}v\textrm{d}u\textrm{d}X
\textrm{d}Y\textrm{d}x\textrm{d}y
\end{eqnarray}
Making a change of variables: $X-x, Y-y \rightarrow X, Y$ in the integral with respect to $X,Y$, we have
\begin{eqnarray}\label{e4}
&&\hat{\rho}_{out}(\hat{X},\hat{Y})
\cr &&~~~~=\frac{1}{\pi}\int W_{in}(X-x,Y-y)G_{\sigma}(x,y)
\cr &&\hskip 0.5in\times \hat{D}(\frac{x+yj}{2})e^{jv[\hat{X}-(X-x)]+ju[\hat{Y}-(Y-y)]}
\cr &&\hskip 0.7in\times \hat{D}^{\dagger}(\frac{x+yj}{2})\textrm{d}v\textrm{d}u\textrm{d}(X-x)
\textrm{d}(Y-y)\textrm{d}x\textrm{d}y
\cr && ~~~~=\frac{1}{\pi}\int G_{\sigma}(x,y)\hat{D}(\frac{x+yj}{2}) W_{in}(X,Y)
\cr &&\hskip 0.4in\times e^{jv(\hat{X}-X)+ju(\hat{Y}-Y)}
\textrm{d}v\textrm{d}u\textrm{d}X
\textrm{d}Y\hat{D}^{\dagger}(\frac{x+yj}{2})\textrm{d}x\textrm{d}y
\cr && ~~~~=\int G_{\sigma}(x,y) \hat{D}(\frac{x+yj}{2})\hat{\rho}_{in}(\hat{X},\hat{Y})\hat{D}^{\dagger}(\frac{x+yj}{2})
\textrm{d}x\textrm{d}y,
\cr &&
\end{eqnarray}
which is just Eq.(\ref{den-inout}).

For the scheme with a parametric amplifier for Bell measurement, we have from Eq.(\ref{W-out-PA})
\begin{equation}
\begin{split}
W_{out}(X,Y)=\int W_{in}(kX-x,\frac{Y}{k}-y)G_{\bar \sigma}(x,y)\textrm{d}x\textrm{d}y.
\end{split}
\end{equation}
where $k\equiv g/G$. Defining $W'_{out}\equiv W_{in}\circ G_{\Bar{\sigma}}$, we have
\begin{equation}\label{e5}
\begin{split}
W_{out}(X,Y)&=W'_{out}(kX,\frac{Y}{k})
\\&=\frac{1}{2\pi}\int \textrm{d}u\bra{kX+u}\hat{\rho}'_{out}\ket{kX-u}e^{-ju\frac{Y}{k}}.
\end{split}
\end{equation}
Making a change of $U=u/k$ in Eq. (\ref{e5}), we have
\begin{equation}
\begin{split}
&W_{out}(X,Y)=\cr
&~~=\frac{k}{2\pi}\int \textrm{d}U\bra{kX+kU}\hat{\rho}'_{out}\ket{kX-kU}e^{-jUY}
\\&~~=\frac{1}{2\pi}\int \textrm{d}U\bra{X+U}\hat{S}(\epsilon)\hat{\rho}'_{out}\hat{S}^{\dagger}(\epsilon)\ket{X-U}e^{-jUY},
\end{split}
\end{equation}
where $\epsilon=-\textrm{ln}(k)=\textrm{ln}(G/g)$. From Eqs.(\ref{e4}) and (\ref{e5}), we have
\begin{equation}
\hat{\rho}'_{out}=\int \hat{D}(\frac{x+yj}{2})\hat{\rho}_{in}\hat{D}^{\dagger}(\frac{x+yj}{2})G_{\sigma}(x,y)\textrm{d}x\textrm{d}y
\end{equation}
Therefore, we obtain Eq.(\ref{rho-out-PA})
\begin{eqnarray}
\hat{\rho}_{out} &=&\int \hat{S}(\epsilon)\hat{D}(\frac{x+yj}{2})\hat{\rho}_{in}\hat{D}^{\dagger}(\frac{x+yj}{2})\cr
&&\hskip 0.7 in \times\hat{S}^{\dagger}(\epsilon)G_{\sigma}(x,y)\textrm{d}x\textrm{d}y.
\end{eqnarray}

\end{document}